\documentclass[a4paper]{jpconf}

\usepackage[square,numbers,merge,comma,sort&compress]{natbib} 
\usepackage{amsmath,amssymb,amsfonts,amsthm,amsbsy}
\usepackage{graphicx}
\graphicspath{{Figures/}}
\usepackage{mathrsfs}
\usepackage{float}
\usepackage[labelsep=colon]{caption}
\usepackage[labelsep=colon,aboveskip=10pt,belowskip=10pt]{subcaption}
\begin{document}

\title{Interaction between superconductors and weak gravitational field}

\author{Antonio Gallerati}

\address{Politecnico di Torino, Dipartimento di Scienza Applicata e Tecnologia, corso Duca degli Abruzzi 24, 10129 Torino, Italy}
\smallskip
\address{Istituto Nazionale di Fisica Nucleare, Sezione di Torino, via Pietro
       Giuria 1, 10125 Torino, Italy}

\ead{antonio.gallerati@polito.it}

\begin{abstract}
We consider the interaction between the Earth's gravitational field and a superconductor in the fluctuation regime. Exploiting the weak field expansion formalism and using time dependent Ginzburg–Landau formulation, we show a possible short-time alteration of the gravitational field in the vicinity of the superconductor.
\end{abstract}

\section{Introduction}
The study of the interaction between superconductors and the gravitational field has received great attention in the last decades, due to its possible applications in both theoretical and applied physics. The seminal paper \cite{DeWitt:1966yi} laid the foundation of the search field, that later led to the Podkletnov and Nieminen pioneering experiment \cite{podkletnov1992possibility}, in which they claimed to have observed a gravitational shielding effect. Since no such effect can occur in the classical framework, several subsequent theoretical papers tried to clarify the possible origin of the gravity/superconductivity interplay in the frame of a quantum field formulation \cite{Modanese:1995tx,Modanese:1996zm,Tajmar:2002gm}.\par
Another step towards the construction of a consistent theory came from the introduction of generalized electric-type fields induced by the presence of a gravitational field \cite{schiff1966gravitation,witteborn1967experimental,witteborn1968experiments}, the generalized field having the form $\mathbf{E}=\mathbf{E}_\textrm{e}+\frac{m}{e}\:\mathbf{E}_\textrm{g}$ and therefore characterized by an electric component $\mathbf{E}_\textrm{e}$ and a gravitational one $\mathbf{E}_\textrm{g}$, $m$ and $e$ being the electron mass and charge. Inspired by these experimental researches, we describe below how the same results can be formally obtained using the gravito-Maxwell formalism. 

\section{Weak field expansion}
Here we consider a nearly flat space-time configuration (weak gravitational field), where the metric $g_{\mu\nu}$ can be expanded as
\begin{equation}
g_{\mu\nu}\:\simeq\:\eta_{\mu\nu}+h_{\mu\nu}\:,
\end{equation}
where $\eta_{\mu\nu}=\mathrm{diag}(-1,+1,+1,+1)$ is the flat Minkowski metric in the mostly plus convention and $h_{\mu\nu}$ is a small perturbation. If we introduce the tensor
\begin{equation}
\bar{h}_{\mu\nu}\:=\:h_{\mu\nu}-\frac12\,\eta_{\mu\nu}\,h\:,
\end{equation}
it can be easily demonstrated that the Einstein equations in the harmonic De Donder gauge $\partial^{\mu}\bar{h}_{\mu\nu}\simeq0$ can be rewritten, in first-order approximation, as \cite{Ummarino:2017bvz,Ummarino:2019cvw}
\begin{equation}
R_{\mu\nu}-\frac12\,g_{\mu\nu}\,R\:=\:\partial^{\rho}\mathscr{G}_{\mu\nu\rho}\:=\:8\pi\mathrm{G}\,T_{\mu\nu}\:,
\end{equation}
having defined the tensor
\begin{equation}
\mathscr{G}_{\mu\nu\rho}\:\equiv\:\partial_{{[}\nu}\bar{h}_{\rho{]}\mu}+\partial^{\sigma}\eta_{\mu{[}\rho}\,\bar{h}_{\nu{]}\sigma}
     \:\simeq\:\partial_{{[}\nu}\bar{h}_{\rho{]}\mu}\:.
\end{equation}

\subsection{Gravito-Maxwell formulation}
We then define the fields
\begin{align}
\mathbf{E}_\textrm{g}=-\frac12\,\mathscr{G}_{00i}=-\frac12\,\partial_{{[}0}\bar{h}_{i{]}0}\,,
\qquad
\mathbf{A}_\textrm{g}=\frac14\,\bar{h}_{0i}\,,
\qquad
\mathbf{B}_\textrm{g}=\frac14\,{\varepsilon_i}^{jk}\,\mathscr{G}_{0jk}\,,
\end{align}
for which we obtain, restoring physical units, the set of equations \cite{Ummarino:2017bvz,Ummarino:2019cvw}:
\begin{equation}
\begin{alignedat}{2}
&\nabla\cdot\mathbf{E}_\text{g}=4\pi\mathrm{G}\,\rho_\text{g}\,,\qquad\; &
&\nabla\cdot\mathbf{B}_\text{g}=0 \,,
\\[2.5\jot]
&\nabla\times\mathbf{E}_\text{g}=-\dfrac{\partial\mathbf{B}_\text{g}}{\partial t}\,,\qquad\; &
&\nabla\times\mathbf{B}_\text{g}=4\pi\mathrm{G}\,\frac{1}{c^2}\,\mathbf{j}_\text{g}
    +\frac{1}{c^2}\,\frac{\partial\mathbf{E}_\text{g}}{\partial t}\,,
\end{alignedat}
\end{equation}
having introduced the mass density \:$\rho_\text{g}\equiv-T_{00}$\: and the mass current density \:$\mathbf{j}_\text{g}\equiv T_{0i}$\,.
The above equations have the same structure of the Maxwell equations, with $\mathbf{E}_\textrm{g}$ and $\mathbf{B}_\textrm{g}$ gravitoelectric and gravitomagnetic field, respectively.

\subsection{Generalized fields and equations}
Now let us consider generalized electric/magnetic fields, scalar and vector potentials, having both electromagnetic and gravitational contributions:
\begin{equation}
\mathbf{E}=\mathbf{E}_\text{e}+\frac{m}{e}\,\mathbf{E}_\text{g}\,,\qquad
\mathbf{B}=\mathbf{B}_\text{e}+\frac{m}{e}\,\mathbf{B}_\text{g}\,,\qquad
V=V_\text{e}+\frac{m}{e}\,V_\text{g}\,,\qquad
\mathbf{A}=\mathbf{A}_\text{e}+\frac{m}{e}\,\mathbf{A}_\text{g}\,,
\label{eq:genfields}
\end{equation}
where $m$ and $e$ identify the mass and electronic charge, respectively.
The generalized Maxwell equations for the above fields then become \cite{Ummarino:2017bvz,Ummarino:2019cvw,Behera:2017voq}:
\begin{equation}
\begin{alignedat}{2}
&\nabla\cdot\mathbf{E}=\left(\frac{1}{\varepsilon_0}+\frac{1}{\varepsilon_\text{g}}\right)\rho\,,\qquad\;&
&\nabla\cdot\mathbf{B}=0 \,,
\\[2.5\jot]
&\nabla\times\mathbf{E}=-\dfrac{\partial\mathbf{B}}{\partial t}\,,\qquad\;&
&\nabla\times\mathbf{B}=(\mu_0+\mu_\text{g})\,\mathbf{j}
    +\frac{1}{c^2}\,\frac{\partial\mathbf{E}}{\partial t}\,,
\end{alignedat}
\end{equation}
where $\varepsilon_0$ and $\mu_0$ are the vacuum electric permittivity and magnetic permeability. In the above expression, $\rho$ and $\mathbf{j}$ are the electric charge density and electric current density, respectively, while the mass density and the mass current density vector have been expressed in terms of the latter as
\begin{equation}
\rho_\text{g}=\frac{m}{e}\,\rho\,,\qquad\quad
\mathbf{j}_\text{g}=\frac{m}{e}\:\mathbf{j}\:,
\end{equation}
while the vacuum \emph{gravitational} permittivity $\varepsilon_\text{g}$ and permeability $\mu_\text{g}$ have the form
\begin{equation}
\varepsilon_\text{g}=\frac{1}{4\pi\mathrm{G}}\,\frac{e^2}{m^2}\:,\qquad\quad
\mu_\text{g}=\frac{4\pi\mathrm{G}}{c^2}\,\frac{m^2}{e^2}\:.
\end{equation}

\section{The quantum model}
Let us now consider a superconductor in the vicinity of its critical temperature. The sample behavior is characterized by thermodynamic fluctuations of the order parameter creating superfluid regions of accelerated electrons, causing in turn an increase of the resistivity for temperatures $T>T_\text{c}$\,.
This regime can be well described using time-dependent Ginzburg-Landau formulation \cite{Cyrot1973Ginzburg} and, if we suppose we deal with sufficiently dirty materials, the effects of the fluctuations can be observed over a sizable range of temperature.\par
The time-dependent Ginzburg-Landau equations characterizing the system, for temperatures larger than $T_\text{c}$, have the gauge-invariant form \cite{hurault1969nonlinear,schmid1969diamagnetic}:
\begin{equation}
\Gamma\left(\hbar\,\partial_t-2\,i\,e\,\phi\right)\psi\:=\,
    \frac{1}{2m}\left(\hbar\,\nabla-2\,i\,e\,\mathbf{A}\right)^2\psi
    +\alpha\,\psi\,.
\end{equation}
We make the following ansatz for the solution
\begin{equation}
\psi(\mathbf{x},t)=f(\mathbf{x},t)\,\exp\big(i\,g(\mathbf{x},t)\big)\,,
\end{equation}
and one then finds for the superfluid speed and the associated current density
\begin{equation}
\mathbf{v}_\text{s}=\frac{1}{m}\left(\hbar\,\nabla g+2\,\frac{e}{c}\,\mathbf{A}\right)\,,
\qquad\;
\mathbf{j}_\text{s}=-2\,\frac{e}{m}\,|\psi|^{2}
    \left(h\,\nabla g+2\,\frac{e}{c}\,\mathbf{A}\right)
    =-2\,e\,f^{2}\,\mathbf{v}_\text{s}\,.
\end{equation}
The latter can be explicitly calculated from \cite{Ummarino:2019cvw}
\begin{equation}
\mathbf{j}_\text{s}(t)=\frac{2\,e^{2}}{m}\,\mathbf{E}\;t\;\frac{\text{k}_\textsc{b\,}T}{8\pi^{3}}
    \!\!\!\!\!\int\limits_{\;0}^{\quad\;+\infty}
        \!\!\!\!dk\;4\pi\,k^{2}\,\left(\alpha+\frac{\hbar^2\,k^2}{2\,m}\right)^{-1}\exp\left(-\frac{2}{\hbar\,\Gamma}\Big(\alpha+\frac{\hbar^2\,k^2\,}{2\,m}\Big)\,t-\frac{4}{3\,\hbar\,\Gamma}\,\frac{e^2}{m}\,E^{2}\,t^3\right),
\end{equation}
having defined the quantities
\begin{equation}
\Delta T=T-T_\text{c}\,,\qquad
\epsilon(T)=\sqrt{\frac{\Delta T}{T_\text{c}}}\,,\qquad
\alpha=\frac{\hbar^{2}}{2\,m\,\xi_0^{\,2}}\,\epsilon(T)\,,\qquad
\Gamma=\frac{\alpha}{\epsilon(T)}\,\frac{\pi}{8\,\text{k}_\textsc{b\,}T_\text{c}}\,,\quad\;\;
\end{equation}
where $\xi_0$ is the BCS coherence length. The potential vector $\mathbf{A}(x,y,z,t)$ is given by:
\begin{equation}
\mathbf{A}(x,y,z,t)=\frac{\mu_0}{4\pi}\int\frac{\mathbf{j}_\text{s}(t)\;\,dx'\,dy'\,dz'}{\sqrt{(x-x')^{2}+(y-y')^{2}+(z-z')^{2}}}\,,
\end{equation}
and the generalized electric field \eqref{eq:genfields} is then written as
\begin{equation}
\mathbf{E}(x,y,z,t)\:=\,-\partial_t\mathbf{A}(x,y,z,t)+\frac{m}{e}\,\mathbf{g}\:=
    \,-\frac{\mu_0}{4\pi}\:\partial_t\,\mathbf{j}_\text{s}(t)\;\mathcal{C}(x,y,z)+\frac{m}{e}\,\mathbf{g}\,,
\end{equation}
featuring the contribution coming from the Earth-surface gravitational field $\mathbf{g}$, while $\mathcal{C}(x,y,z)$ is a geometrical factor whose expression depends on the shape of the superconducting sample.

\section{Experimental predictions}
Let us now consider the case of a superconducting sample, at a temperature very close to $T_\text{c}$, that is put it in the normal state with a weak magnetic field. The latter is then removed at the time $t=0$, so that the system enters the superconducting state. Using the described quantum model, we can calculate the variation of the gravitational field in the vicinity of the sample, in the fluctuation regime.\par
\begin{figure}[htp]
\centering
\captionsetup[figure]{skip=6pt,belowskip=-5pt,font=footnotesize,labelfont=footnotesize,labelfont=bf,margin=0.25cm,format=hang}
%
\begin{minipage}{.485\textwidth}
\centering
\includegraphics[width=\textwidth,keepaspectratio]{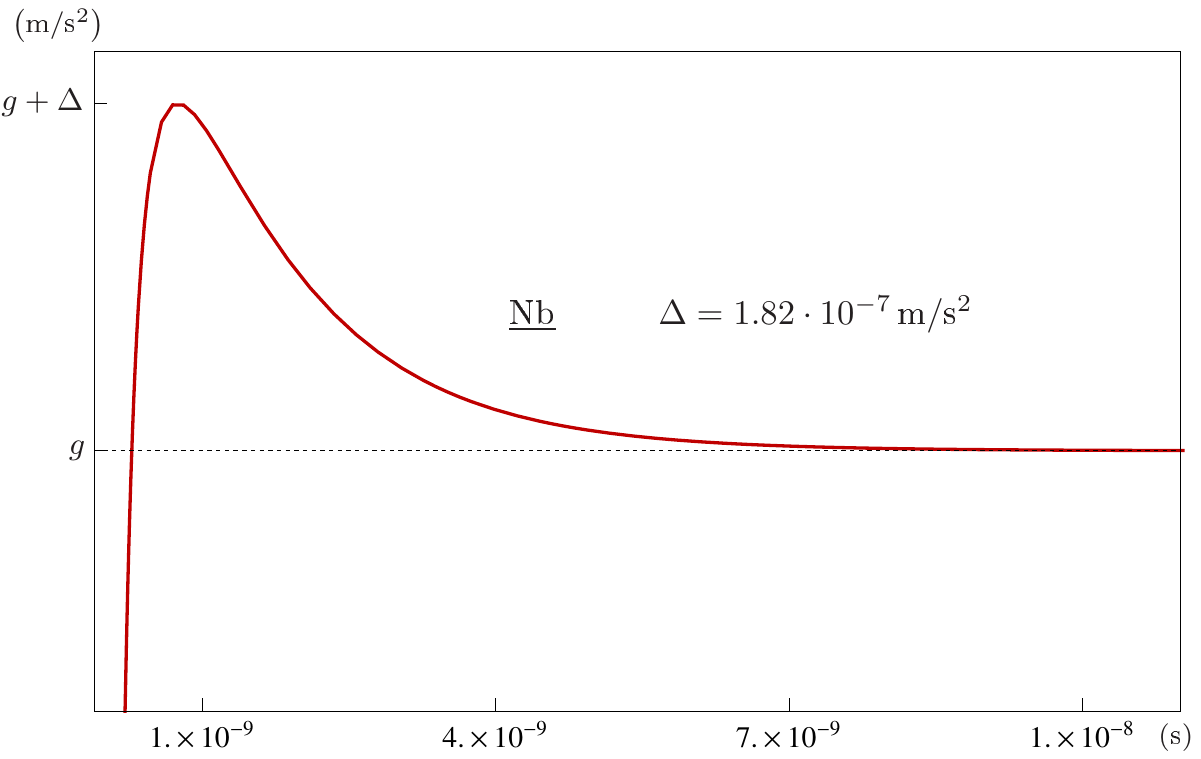}
\captionof{figure}{The gravitational field variation as a function of time for a Nb sample measured along the axis of the disk at fixed distance $d$ above the base surface.}
\label{fig:Nbtime}
\end{minipage}
\hfill
\begin{minipage}{.49\textwidth}%
\centering
\includegraphics[width=\textwidth,keepaspectratio]{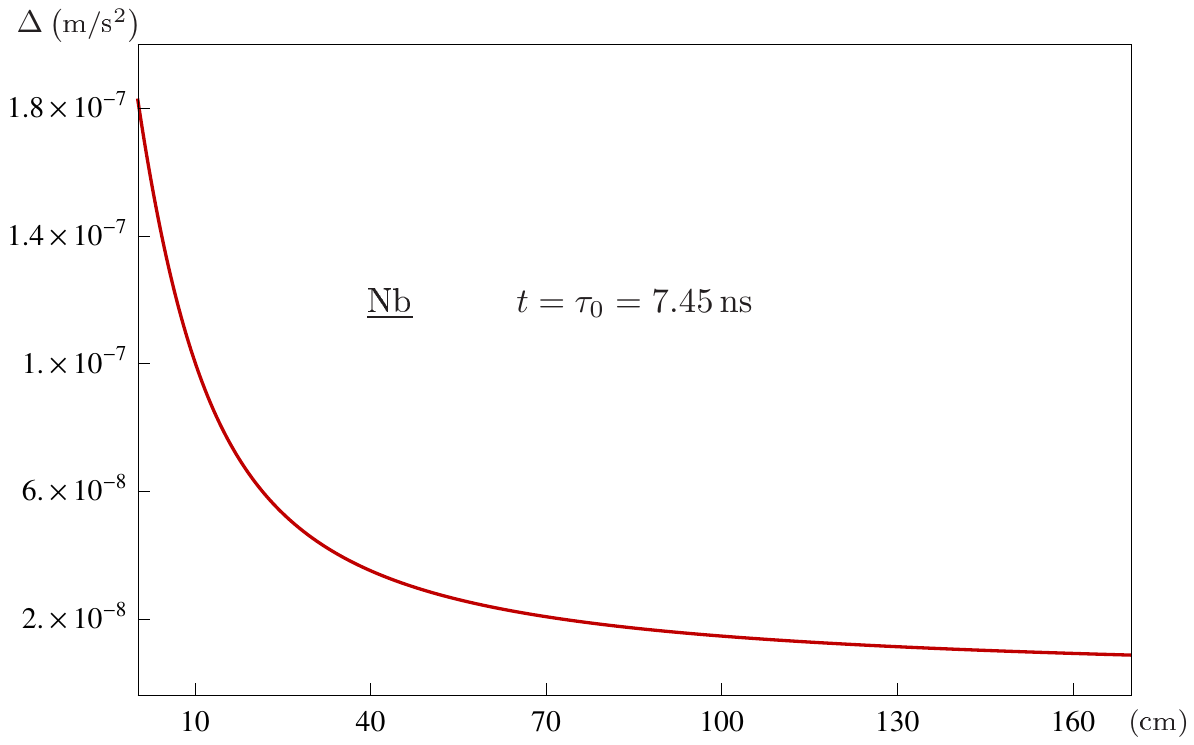}
\captionof{figure}{The gravitational field variation as a function of distance from the base surface for a Nb sample, measured at the fixed time $t=\tau_0$.}
\label{fig:Nbdistance}
\end{minipage}
%
\end{figure}

In particular, let us consider a superconducting disk with bases parallel to the ground. In Figure \ref{fig:Nbtime} is plotted the variation of the gravitational field as a function of time, measured along the axis of the disk at fixed distance $d=0.25\,\mathrm{cm}$ above the base surface, for a Nb sample (low-$T_\text{c}$ superconductor, \,$\xi_0=39\,\mathrm{nm}$, \,$T_\text{c}=9.250\,\mathrm{K}$, \,$\Delta T=10^{-3}\,\mathrm{K}$ \,\cite{poole1999handbook}) having radius $R=15\,\mathrm{cm}$ and thickness $h=2\,\mathrm{cm}$. We can appreciate that the gravitational field is initially reduced with respect to its unperturbed value, then subsequently increases up to a maximum value for $t=\tau_{0}$ and finally relaxes to the standard external value.
In Figure \ref{fig:Nbdistance} we show the field variation as a function of distance from the base surface, measured along the axis of the disk at the fixed time $t=\tau_0=7.45\,\mathrm{ns}$ that maximizes the effect.\par
\begin{figure}[htp]
\centering
\captionsetup[figure]{skip=6pt,belowskip=-5pt,font=footnotesize,labelfont=footnotesize,labelfont=bf,margin=0.25cm,format=hang}
%
\begin{minipage}{.485\textwidth}
\centering
\includegraphics[width=\textwidth,keepaspectratio]{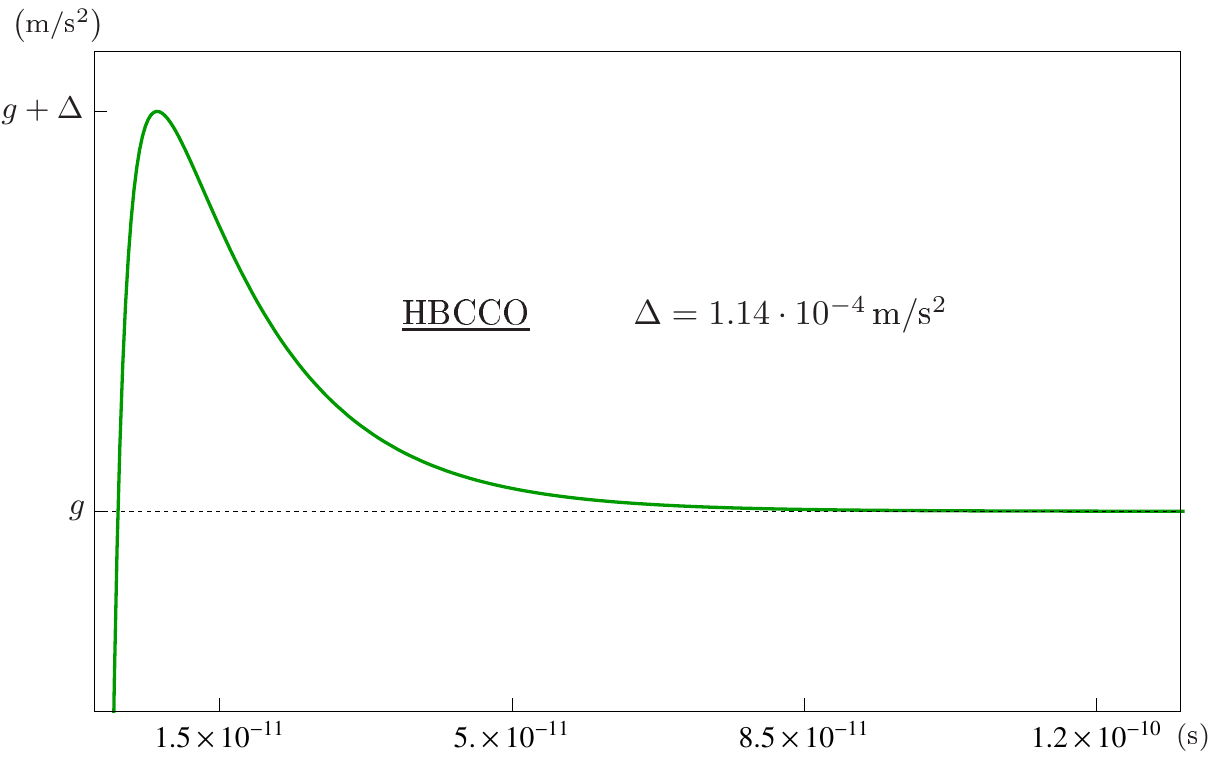}
\captionof{figure}{The gravitational field variation as a function of time for a HBCCO sample measured along the axis of the disk at fixed distance $d$ above the base surface.}
\label{fig:HBCCOtime}
\end{minipage}
\hfill
\begin{minipage}{.49\textwidth}%
\centering
\includegraphics[width=\textwidth,keepaspectratio]{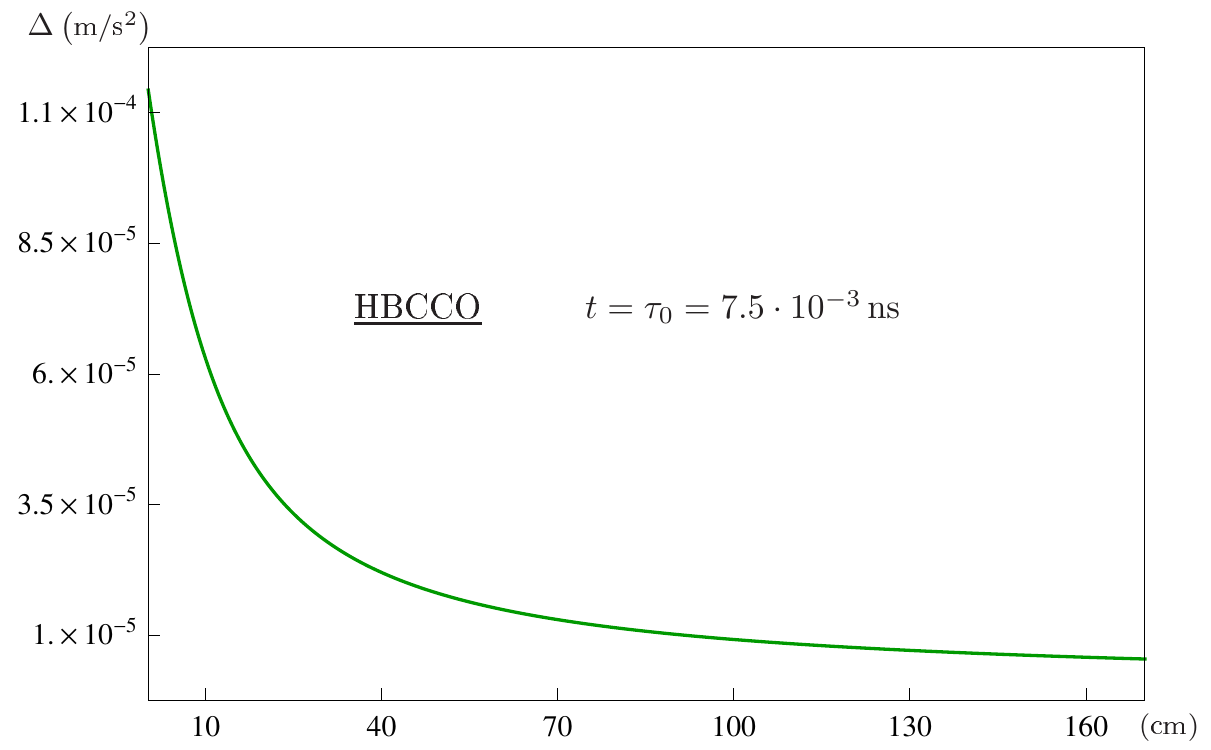}
\captionof{figure}{The gravitational field variation as a function of distance from the base surface for a HBCCO sample, measured at the fixed time $t=\tau_0$.}
\label{fig:HBCCOdistance}
\end{minipage}
%
\end{figure}

In Figures \ref{fig:HBCCOtime} and \ref{fig:HBCCOdistance} the same calculations are performed using a HgBaCaCuO (HBCCO) sample, an high-$T_\text{c}$ superconductor ($\xi_0=230\,\mathrm{nm}$, \,$T_\text{c}=126\,\mathrm{K}$, \,$\Delta T=0.1\,\mathrm{K}$ \,\cite{poole1999handbook}).\par
It is easily shown that the maximum value $\Delta$ for the variation of the external field is proportional to ${\xi_0}^{-1}$, implying a larger effect in high-$T_\text{c}$ superconductors, having the latter small coherence length.
It is also possible to demonstrate that $\tau_{0}\propto (T-T_\text{c})^{-1}$, that in turn means that the time range in which the phenomenon takes place can be extended if the system is very close to its critical temperature.

\section{Conclusions and future developments}
As can be seen from the results obtained, the field variation is in principle perceptible (especially in high-$T_\text{c}$ superconductors), while the very short time intervals in which the effect occurs complicate direct measurements. In order to obtain non negligible experimental evidence of gravitational perturbations in workable time scales, a careful choice of parameters must be made.
First of all, a large superconducting sample of dirty material is needed, so that the effects of fluctuations can be enhanced over a wider temperature range. Then, the best option currently is to choose an high-$T_\text{c}$ superconductor (short coherence length increases the intensity of the phenomenon) at a temperature very close to $T_\text{c}$ (increase in the time interval where the effect occurs).\par
Possible future developments of the described formalism derive from the application to different physical situation where generalized electric-magnetic fields of the form \eqref{eq:genfields} are induced by the presence of a weak gravitational field. An example of application to the Josephson junction physics of superconductors can be found in \cite{Ummarino2020josephson}.

\bibliographystyle{iopart-num}
\bibliography{bibliografia} 

\end{document}